\documentclass[floatfix,aps,pra,twocolumn,showpacs,superscriptaddress,10pt]{revtex4-1}
\usepackage[utf8]{inputenc}
\usepackage[T1]{fontenc}
\usepackage{color}
\usepackage{float}
\usepackage{mathrsfs}
\usepackage{amstext}
\usepackage{amsmath}
\usepackage{graphicx}
\usepackage{comment}
\usepackage{dsfont}

\usepackage{lipsum}

\usepackage{babel}
\begin{document}

\preprint{APS/123-QED}

\title{Sensitivity of electromagnetically induced transparency to light-mediated interactions}

\author{M. H. Oliveira}
\email{murilo.oliveira@df.ufscar.br}
\affiliation{Departamento de Física, Universidade Federal de São Carlos, P.O. Box
676, 13565-905, São Carlos, São Paulo, Brazil.}

\author{C. E. Máximo}
\affiliation{Departamento de Física, Universidade Federal de São Carlos, P.O. Box
676, 13565-905, São Carlos, São Paulo, Brazil.}

\author{C. J. Villas-Boas}
\affiliation{Departamento de Física, Universidade Federal de São Carlos, P.O. Box
676, 13565-905, São Carlos, São Paulo, Brazil.}

\begin{abstract}
Here we present a microscopic model that describes the Electromagnetically Induced Transparency (EIT) phenomenon in the multiple scattering regime. We consider an ensemble of cold three-level atoms, in a $\Lambda$ configuration, scattering a probe and a control field to the vacuum modes of the electromagnetic field. By first considering a scalar description of the scattering, we show that the light-mediated long-range interactions that emerge between the dipoles narrow the EIT transparency window for increasing densities and sample sizes. For a vectorial description, we demonstrate that near-field interacting terms can critically affect the atomic population transfer in the Stimulated Raman Adiabatic Passage (STIRAP). This result points out that standard STIRAP-based quantum memories in cold atomic ensembles would not reach high enough efficiencies for quantum information processing applications even in dilute regimes.

\end{abstract}

\maketitle

\section{Introduction}
Electromagnetically Induced Transparency (EIT) \cite{Harris1990} is a quantum interference phenomenon in which an initially opaque ensemble of three-level atoms becomes transparent to a probe field due to the influence of a second field, known as the control field. The existence of a dark state in the atomic system is the reason why the absorption ceases when the probe field is tuned at resonance with a given atomic transition.   In particular, when probe and control fields are of the same magnitude, such a dark state becomes a superposition of two atomic ground states, which gives origin to a Coherent Population Trapping (CPT) in the steady-state regime \cite{alzetta1976,arimondo1976}. 

EIT and CPT phenomena have been receiving substantial attention thanks to their vast list of applications \cite{ImamogluRev2005}. For example, EIT is useful for the reduction of the group velocity of a light pulse which propagates through an atomic medium \cite{SlowLightNature}, for the narrowing of the transmission linewidth of optical cavities \cite{WangNarrowing, Lukin}, and for quantum memory implementations, where photonic states can be mapped and stored in single atoms trapped inside optical cavities \cite{Specht2011,MemoryCelso}, or in an atomic ensemble \cite{LiuMemory}. For the latter, it has been theoretically demonstrated that the efficiency of quantum memory devices increases with the sample optical thickness \cite{gorshkov2007universal,ma2017optical}. Indeed high efficiency ($>90\%$) in retrieving quantum information has been achieved in cold-atom platforms only for high optical thickness \cite{hsiao2018,vernaz2018,wang2019}, a regime where multiple scattering of light becomes relevant. This leads us to the unexplored question of how coherent collective scattering of light would affect EIT and CPT transparency windows and, consequently, all corresponding applications. The purpose of the present letter is to shine a light on this question.

In light scattering by cold atoms, effective light-mediated interactions emerge between all scatterers \cite{scully2010,kaiser2011,kaiser2013} owing to a strong suppression of the Doppler broadening by laser cooling techniques. Such optical dipole-dipole interactions give origin to several collective effects, as superradiance \cite{felinto2014,guerin2016,sokolov2016}, subradiance \cite{kaiser2016,bachelard2018,guerin2018,rey2019,rey2020}, coherent backscattering of light \cite{kaiser1999,kaiser2014} and cooperative Lamb shifts \cite{rey2016,sokolov2016}, and cover long ranges in a similar fashion as Rydberg atomic interactions. Yet Rydberg interactions are Hamiltonian interactions that do not depend on light scattering to remain active, being even able to totally destroy EIT \cite{adams2008,adams2010,fleischhauer2011,zhang2018Rydberg}. On the other hand, light-mediated interactions are a consequence of the collective scattering of light in the atomic sample, so any population of a dark state naturally reduces the atomic cooperation in some level. This is the reason why purely optical interactions should induce more subtle modifications in transparency windows, and it is not clear yet to which extent the EIT/CPT applications are affected.
\begin{figure}[ht]
\begin{centering}
\includegraphics[width=1.0\columnwidth]{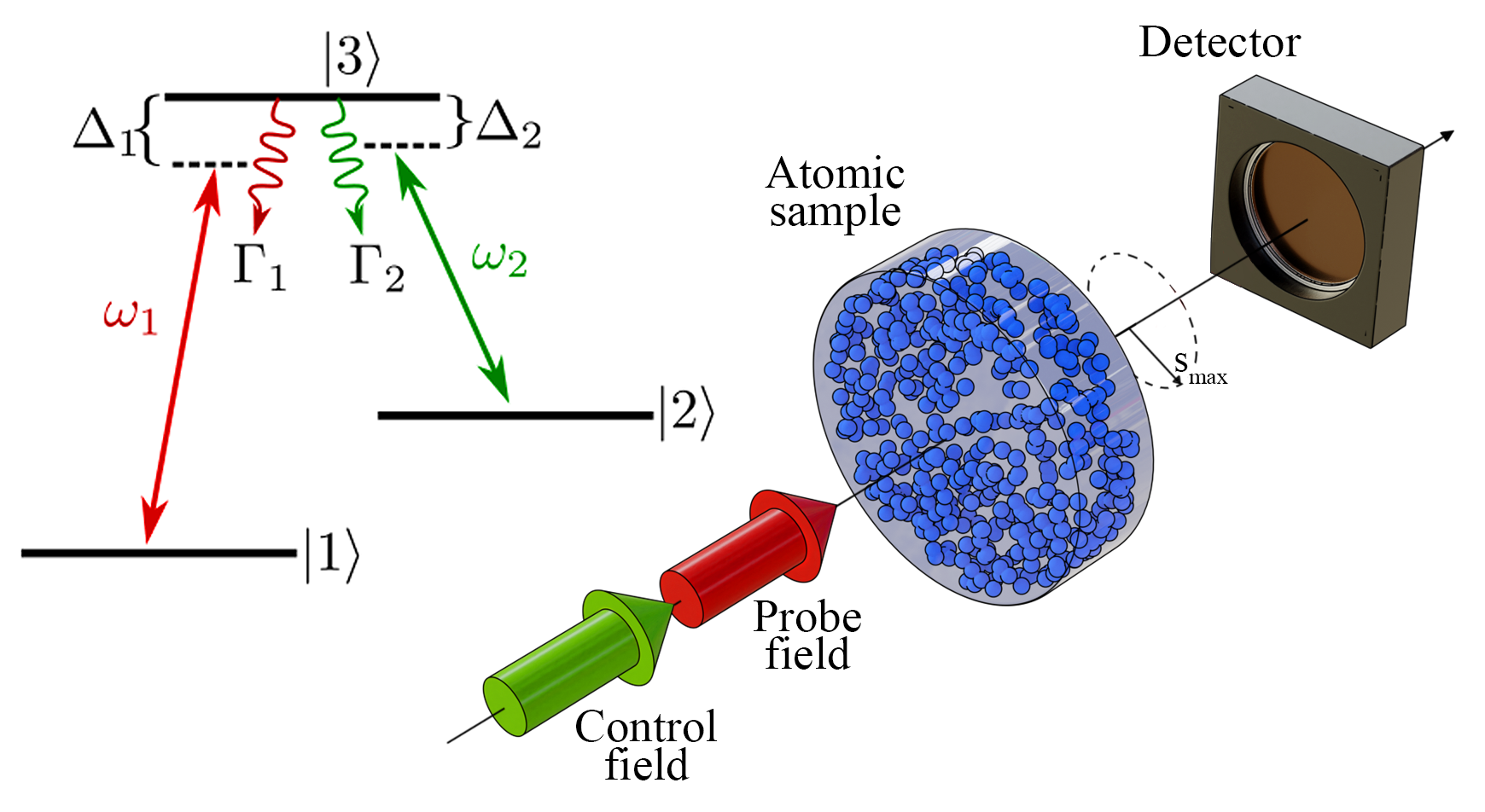}
\par\end{centering}
\caption{A cylindrical and homogeneous cold cloud of $N$ disordered three-level atoms scatters probe and control fields to the free space. The atomic levels are in a $\Lambda$ configuration, as schematically shown on the left. The radius $R$ of the cylindrical surface is much larger than its thickness $L$, measuring the probe field transmission in a disk of radius $s_{\text{max}}<R$.}
\label{fig:1}
\end{figure}

In order to detect collective modifications on EIT/CPT phenomena, here we derive a microscopic coupled-dipole model that describes the multiple scattering of probe and control fields by cold ensembles of three-level atoms. Such a model represents a significant extension of the linear-optics coupled-dipole model \cite{scully2010,kaiser2011,kaiser2013,sokolov2019}. In the limit where polarization effects can be neglected, namely the dilute regime, we show that light-mediated interactions narrow the width of the transparency window at its Full Width at Half Maximum (FWHM). Furthermore, we analyse the Stimulated Raman Adiabatic Passage (STIRAP) to get a prospect of the multiple scattering effect on the fore-mentioned quantum memory applications. This EIT-based dynamical technique is a fundamental ingredient in writing and retrieving protocols for quantum memories \cite{Bergmann2019}.  While considering a complete vectorial description, we demonstrate that the efficiency of such a STIRAP process, in which the population is exchanged between the two ground states of the $\Lambda$-system, is substantially reduced by collective scattering of light even in the dilute regime. As our model combines long-range interactions mediated by light with two distinct incident fields, it opens a new route for the study of collective effects in nonlinear optics.

\section{Microscopic model}
We consider an ensemble of $N$ point-like three-level atoms at zero temperature, with random positions $\mathbf{r}_{j}=\left(x_{j},y_{j},z_{j}\right)$, for $j=1,...,N$, decaying to the vacuum modes of the radiation field. Their $\Lambda$ energy-level scheme is composed of two ground states, $\vert 1 \rangle$ and $\vert 2 \rangle$, and one excited state $\vert 3 \rangle$, as represented in Fig. \ref{fig:1}.
A probe field, of angular frequency $\omega_1$ and Rabi frequency $\Omega_1$, pumps the transition $\vert 1 \rangle \leftrightarrow \vert 3 \rangle$, of frequency $\omega_{31}$, whereas a control field, of angular frequency $\omega_2$ and Rabi frequency $\Omega_2$, drives the transition $\vert 2\rangle \leftrightarrow \vert 3 \rangle$, of frequency $\omega_{32}$. Both fields are plane waves propagating in the direction of the wave vectors $\mathbf{k}_{1}$ and $\mathbf{k}_{2}$, respectively. The light-matter Hamiltonian that describes this system reads:
\begin{multline}
\hat{H}=\sum_{j=1}^{N}\sum_{n=1}^{3}\omega_{nn}\hat{\sigma}_{nn}^{j}+\sum_{\mathbf{k},\mathbf{s}}\nu_{\mathbf{k}}\left(\hat{a}_{\mathbf{k},\mathbf{s}}^{\dagger}\hat{a}_{\mathbf{k},\mathbf{s}}+\frac{1}{2}\right)\\+\sum_{j=1}^{N}\sum_{n=1}^{2}\sum_{\mathbf{k},\mathbf{s}}g_{\mathbf{k},\mathbf{s}}^{\left(n\right)}\left(\hat{\sigma}_{n3}^{j}+\hat{\sigma}_{3n}^{j}\right)\left(\hat{a}_{\mathbf{k},\mathbf{s}}^{\dagger}e^{-i\mathbf{k}\cdot\mathbf{r}{}_{j}}+\hat{a}_{\mathbf{k},\mathbf{s}}e^{i\mathbf{k}\cdot\mathbf{r}{}_{j}}\right)\\+\frac{1}{2}\sum_{j=1}^{N}\sum_{n=1}^{2}\Omega_{n}\left(\hat{\sigma}_{n3}^{j}e^{i\omega_{n}t-i\mathbf{k}_{n}\cdot\mathbf{r}{}_{j}}+\hat{\sigma}_{3n}^{j}e^{-i\omega_{n}t+i\mathbf{k}_{n}\cdot\mathbf{r}{}_{j}}\right),\label{H}
\end{multline}
where $\hat{\sigma}_{n3}^{j}=\vert n \rangle\langle 3 \vert^{j}$ ($\hat{\sigma}_{3n}^{j}=\vert 3 \rangle\langle n \vert^{j}$) are the lowering (raising) atomic operators, and $\hat{\sigma}_{nn}^{j}=\vert n \rangle\langle n \vert^{j}$ are the atomic population operators. Each vacuum mode is characterized by its wave vector $\mathbf{k}$, polarization vector $\mathbf{s}$, and angular frequency $\nu_{\mathbf{k}}$, where $\hat{a}_{\mathbf{k}}^{\dagger}$ ($\hat{a}_{\mathbf{k}}$) is the corresponding creation (annihilation) operator. The exchange of photons between the atoms and the environment takes place with coupling strength $g_{\mathbf{k},\mathbf{s}}^{\left(n\right)}=\mathbf{s}\cdot\mathbf{d}_{n}\sqrt{\nu_{\mathbf{k}}/2\epsilon_{0}V_{\mathbf{k}}}$, for $\mathbf{d}_{n}$, $\epsilon_{0}$ and $V_{\mathbf{k}}$, respectively, the dipole matrix elements of the transitions, the vacuum permittivity and the mode volumes. 

The explicit time dependence on the laser terms, appearing in the last line of  Eq.\eqref{H}, can be removed by applying two consecutive unitary transformations: $e^{-i H_0 t}$, with $H_0$ representing the free energy terms, and $e^{-i\left[\sum_{j}\left(\Delta_{1}-\Delta_{2}\right)\sigma_{22}^{j}+\Delta_{1}\sigma_{33}^{j}\right]t}$, where we define the detuning $\Delta_{n}=\omega_{n}- \omega_{3n}$. Then the following Hamiltonian is obtained:
\begin{eqnarray}
\mathcal{\hat{H}}&=&\sum_{j=1}^{N}\left[\left(\Delta_{1}-\Delta_{2}\right)\hat{\sigma}_{22}^{j}+\Delta_{1}\hat{\sigma}_{33}^{j}\right]\nonumber\\&&+\frac{1}{2}\sum_{j=1}^{N}\sum_{n=1}^{2}\Omega_{l}\left(\hat{\sigma}_{n3}^{j}e^{-i\mathbf{k}_{n}\cdot\mathbf{r}{}_{j}}+\hat{\sigma}_{3n}^{j}e^{i\mathbf{k}_{n}\cdot\mathbf{r}{}_{j}}\right)\nonumber\\&&+\sum_{j=1}^{N}\sum_{n=1}^{2}\left(\hat{\sigma}_{n3}^{j}e^{-i\omega_{n}t}+\hat{\sigma}_{3n}^{j}e^{i\omega_{n}t}\right)\nonumber\\&&\times\sum_{\mathbf{k},\mathbf{s}}g_{\mathbf{k},\mathbf{s}}^{\left(n\right)}\left(\hat{a}_{\mathbf{k},\mathbf{s}}^{\dagger}e^{i\nu_{\mathbf{k}}t-i\mathbf{k}\cdot\mathbf{r}{}_{j}}+\hat{a}_{\mathbf{k},\mathbf{s}}e^{-i\nu_{\mathbf{k}}t+i\mathbf{k}\cdot\mathbf{r}{}_{j}}\right). \label{Ht}
\end{eqnarray}
Note that the time dependence was transferred to the interaction between atoms and vacuum modes, which is the required expression for the next steps.

In order to obtain the expectation values dynamics for the atomic operators, we first evolve atom and photon operators in the Heisenberg representation, according to the Hamiltonian \eqref{Ht}. Then we formally solve the equations for the photon operators and substitute their solutions in the dipole equations \cite{kaiser2011,bachelard2015,rey2016}. Neglecting fasting oscillating terms in the Markov approximation, we then obtain the following reduced dynamics:
\begin{eqnarray}
\frac{d\hat{\sigma}_{nn}^{j}}{dt}&=&\frac{\Gamma_{n}}{2}\hat{\sigma}_{33}^{j}+\frac{1}{2}\left(\hat{\sigma}_{3n}^{j}\mathcal{\hat{F}}_{n}^{j}+\text{H.c.}\right),\label{sig11}\\
\frac{d\hat{\sigma}_{n3}^{j}}{dt}&=&-\left(\frac{\Gamma}{2}+i\Delta_{n}\right)\hat{\sigma}_{n3}^{j}\nonumber\\&&-\frac{1}{2}\left[\left(\hat{\sigma}_{nn}^{j}-\hat{\sigma}_{33}^{j}\right)\mathcal{\hat{F}}_{n}^{j}+\hat{\sigma}_{nm}^{j}\mathcal{\hat{F}}_{m}^{j}\right], \\
\frac{d\hat{\sigma}_{12}^{j}}{dt}&=&i\left(\Delta_{2}-\Delta_{1}\right)\hat{\sigma}_{12}^{j}\nonumber\\&&+\frac{1}{2}\left[\hat{\sigma}_{32}^{j}\mathcal{\hat{F}}_{1}^{j}+\hat{\sigma}_{13}^{j}\left(\mathcal{\hat{F}}_{2}^{j}\right)^{\dagger}\right], \label{sig12}
\end{eqnarray}
for $m,n=1,2$ with $m\neq n$, where we define the effective field operators 
\begin{equation}
\mathcal{\hat{F}}_{n}^{j} \equiv i\hat{\mathds{1}}\Omega_{n}e^{i\mathbf{k}_{n}\cdot\mathbf{r}{}_{j}}+\sum_{l\neq j}\hat{\sigma}_{n3}^{l}G_{n}^{jl} +\text{Noise},\label{field}
\end{equation}
and the effective light-mediated interactions
\begin{equation}
G_{n}^{jl}\equiv2\int_{0}^{\infty}d\tau e^{i\omega_{n}\tau}\sum_{\mathbf{k},\mathbf{s}}\Bigl|g_{\mathbf{k},\mathbf{s}}^{\left(n\right)}\Bigr|^{2}\left(e^{-i\nu_{\mathbf{k}}\tau+i\mathbf{k}\cdot\mathbf{r}_{jl}}-c.c.\right). \label{kernel}
\end{equation}
In Eqs.\eqref{sig11}-\eqref{sig12}, the decay rates that determine the time scale of the problem are $\Gamma_n \equiv \text{Re}\left(G_{n}^{jj}\right)$, with $\Gamma \equiv \Gamma_1+\Gamma_2$. Whereas in Eq.~\eqref{kernel}, we have defined $\mathbf{r}_{jl} \equiv \mathbf{r}_{j}-\mathbf{r}_{l}$ as the relative position between atoms $j$ and $l$. Note that the upper limit of the time integral in Eq.~\eqref{kernel} is now the infinity since the vacuum modes dynamics are much faster then the population dynamics.

Going to the spherically symmetric continuous integration
\begin{equation}
\sum_{\mathbf{k}}\rightarrow\frac{V}{\left(2\pi\right)^{3}}\int_{0}^{\infty}k^{2}dk\int_{0}^{\pi}\sin\theta d\theta\int_{0}^{2\pi}d\phi, \label{kvec}
\end{equation} 
which covers all possible wave vectors $\mathbf{k}$ \cite{scully2010,kaiser2011,kaiser2013}, we end up with effective long-range interactions, 
\begin{multline}
G_{n}^{jl}=\frac{3}{2}\frac{\Gamma_n e^{ik_{n}r_{jl}}}{ik_{n}r_{jl}}\left[1+\frac{i}{k_{n}r_{jl}}-\frac{1}{k_{n}^{2}r_{jl}^{2}}\right. \\
\left.-\left(\frac{z_{jl}}{r_{jl}}\right)^2\left(1+\frac{3i}{k_{n}r_{jl}}-\frac{3}{k_{n}^{2}r_{jl}^{2}}\right)\right], \label{G}
\end{multline}
that decay with the euclidean distance $r_{jl}=\vert \mathbf{r}_{jl} \vert$ between atomic pairs. Here, we have defined $z_{jl}=z_j-z_l$, where $z_j$ is the position of the \textit{j-th} atom along the cylinder's longitudinal axis. For the derivation of the interactions $G_{n}^{jl}$, one takes into account the polarization of all vacuum modes in the radiation-matter Hamiltonian \cite{rey2016}. This is the reason of the name ``vectorial'' for the corresponding scattering model. However, dilute atomic clouds ($\rho/k_1^3 < 0.01$) are well described by the scalar approximation \cite{skipetrov2014,kaiser2011}
\begin{equation}
    G_{n}^{jl} \approx \frac{\Gamma_{n} e^{ik_nr_{jl}}}{ik_nr_{jl}},
\end{equation}
which can be obtained by not taking into account the polarization vectors ${\mathbf{s}}$ in the initial Hamiltonian $\hat{H}$. In this work, we consider the scalar model for the calculation of the transparency window, while both models are compared in the collective STIRAP analysis. 

Returning to Eqs. (3)-(5), we now obtain the dynamical equations for the expectation values $\langle\:\cdot\:\rangle$ of the dipole operators $\hat{\sigma}_{nn}^{j}$. We consider that the total density matrix of the system can be approximated by the tensor product between the individual density matrices of each subsystem $\left(\hat{\rho}=\hat{\rho}^{1}\otimes\hat{\rho}^{2}\otimes...\otimes\hat{\rho}^{N}\right)$, which in our case is a good approximation for incident fields such that $\Omega_n< \Gamma_n$ \cite{kramer2015}. As a result, we can neglect the correlations between different dipoles $\langle\hat{\sigma}_{nm}^{j}\hat{\sigma}_{n'm'}^{l}\rangle\approx \langle\hat{\sigma}_{nm}^{j}\rangle\langle\hat{\sigma}_{n'm'}^{l}\rangle$, for $j\neq l$, while still keeping single-atom correlations $\langle\hat{\sigma}_{nm}^{j}\hat{\sigma}_{n'm'}^{j}\rangle$. This type of semiclassical approximation leads us to the following system of equations: 
\begin{eqnarray}
\frac{d\bigl\langle\hat{\sigma}_{nn}^{j}\bigr\rangle}{dt}&=&\frac{\Gamma_{n}}{2}\bigl\langle\hat{\sigma}_{33}^{j}\bigr\rangle+\text{Re}\left(\bigl\langle\hat{\sigma}_{3n}^{j}\bigr\rangle\bigl\langle\mathcal{\hat{F}}_{n}^{j}\bigr\rangle\right),\label{<sig11>}\\
\frac{d\bigl\langle\hat{\sigma}_{n3}^{j}\bigr\rangle}{dt}&=&-\left(\frac{\Gamma}{2}+i\Delta_{n}\right)\bigl\langle\hat{\sigma}_{n3}^{j}\bigr\rangle-\frac{1}{2}\bigl\langle\hat{\sigma}_{nm}^{j}\bigr\rangle\bigl\langle\mathcal{\hat{F}}_{m}^{j}\bigr\rangle \nonumber \\ &&-\frac{1}{2}\left(\bigl\langle\hat{\sigma}_{nn}^{j}\bigr\rangle-\bigl\langle\hat{\sigma}_{33}^{j}\bigr\rangle\right)\bigl\langle\mathcal{\hat{F}}_{n}^{j}\bigr\rangle,\label{<sig23>}\\
\frac{d\bigl\langle\hat{\sigma}_{12}^{j}\bigr\rangle}{dt}&=&i\left(\Delta_{2}-\Delta_{1}\right)\bigl\langle\hat{\sigma}_{12}^{j}\bigr\rangle \nonumber\\
&&+\frac{1}{2}\left(\bigl\langle\hat{\sigma}_{32}^{j}\bigr\rangle\bigl\langle\mathcal{\hat{F}}_{1}^{j}\bigr\rangle+\bigl\langle\hat{\sigma}_{13}^{j}\bigr\rangle\bigl\langle\mathcal{\hat{F}}_{2}^{j}\bigr\rangle^{*}\right), \label{<sig12>}
\end{eqnarray}
for $m,n=1,2$ with $m\neq n$.
Note that the contribution from noise operators disappears in the semiclassical approximation since $\bigl\langle\hat{a}_{\mathbf{k},\mathbf{s}}\left(0\right)\bigr\rangle=0$ for a white noise reservoir. As another consequence of the semiclassical approximation, the expectation values
\begin{equation}
\bigl\langle\mathcal{\hat{F}}_{n}^{j}\bigr\rangle=i\Omega_{n}e^{i\mathbf{k}_{n}\cdot\mathbf{r}{}_{j}}+\sum_{l\neq j} G_{n}^{jl} \bigl\langle\hat{\sigma}_{n3}^{l}\bigr\rangle \label{<field>}
\end{equation}
act like mean fields exciting the atomic transitions. The first contribution comes from the incident fields, while the second contribution represents the influence of all the other atoms on a given atom $j$, described by the light-mediated interactions from Eq. \eqref{G} \cite{piovella2014,rey2016}. 

Finally, we point out that the optical interactions $G_{n}^{jl}$ are the main ingredient that distinguish  Eqs. \eqref{<sig11>}-\eqref{<sig12>} from those that describe independent three-level atoms. Note that it couples the transition $\vert 1 \rangle \leftrightarrow \vert 3 \rangle$ of a given atom to the transition $\vert 2 \rangle \leftrightarrow \vert 3 \rangle$ from another atom. An effect neglected in single-scattering models but that here is essential for our many-atom analyses, as can be seen in the following.

\section{Collective Transparency Window}
The procedure adopted in most experiments to investigate the transparency properties of a medium relies on light transmission measurements \cite{Harris1990,ImamogluRev2005,adams2008,adams2010,fleischhauer2011}. In this context, the total scalar electric field operator \cite{courteille2011,kaiser2011,kaiser2013},
\begin{equation}
\hat{E}\left(\mathbf{r},t\right)=\hat{\mathds{1}} E_{1}e^{i\mathbf{k}_{1}\cdot\mathbf{r}}-\frac{\Gamma_{1}}{2d_{1}}\sum_{j=1}^{N}\hat{\sigma}_{13}^{j}\left(t\right)\frac{e^{ik_{1}\left|\mathbf{r}-\mathbf{r}_{j}\right|}}{k_{1}\left|\mathbf{r}-\mathbf{r}_{j}\right|},
\label{E}
\end{equation}
is investigated around the transition frequency $\omega_{31}$, within the spectral range where the contribution of the control field can be neglected. In Eq.\eqref{E}, the probe field, of amplitude $E_{1} \equiv \Omega_{1}/d_{1}$, interferes with its associated scattered field, and generates an intensity profile $I\left(\mathbf{r},t\right) \propto \bigl\langle\hat{E}^{\dagger}\left(\mathbf{r},t\right)\hat{E}\left(\mathbf{r},t\right)\bigr\rangle$ over the whole three-dimensional space. The symbol $\hat{\mathds{1}}$ represents the identity operator of the atomic Hilbert space. To keep the consistence of our procedure, we also consider the semiclassical approximation for the intensity:
\begin{equation}
I\left(\mathbf{r},t\right)  \propto  \bigl|\bigl\langle\hat{E}\left(\mathbf{r},t\right)\bigr\rangle\bigr|^{2} +  \frac{\Gamma_{1}^{2}}{4d_{1}^{2}}\sum_{j=1}^{N}\frac{\bigl\langle\sigma_{33}^{j}\left(t\right)\bigr\rangle-\bigl|\bigl\langle\sigma_{13}^{j}\left(t\right)\bigr\rangle\bigr|^{2}}{k_{1}^{2}\left|\mathbf{r}-\mathbf{r}_{j}\right|^{2}},\label{EE}
\end{equation}
where only quantum correlations between different dipoles were again neglected. 

In our simulations, probe and control fields propagate along the longitudinal axis of a cylinder with thickness $L$ and radius $R$, where the random positions of the atoms are homogeneously distributed with a given average density $\rho = N/L\pi R^2$ (see Fig. \ref{fig:1}) \cite{skipetrov2019}. We then obtain the transmission  $T$ by numerically integrating the steady-state solution of Eq.\eqref{EE} in a disk of area $\mathcal{A}=\pi s_{max}^2$, at the observation point $z_0$, for a radius of integration $s_{max}$. The result is divided by the incident power $E_1^2 \pi s_{max}^{2}$, as in the following:
\begin{equation}
T=\frac{1}{E_1^2 \pi s_{max}^{2}}\int d^{2}sI\left(\mathbf{s},z_{0},t\rightarrow\infty\right).
\end{equation}
To minimize the power losses by diffraction effects around the cylinder edges, we chose $R \gg L$ and $s_{\textrm{max}}<R$ \cite{javanainen2017}. Given that most experiments are carried out in the optical regime and the specific range of parameters we are adopting, the fundamental lengths $L$ and $R$ would be in the scale of $\mu$m. We calculate many realizations of the EIT transmission spectrum as a function of the probe field detuning $\Delta_1$, keeping the radius $R$ constant for different values of the density $\rho$ and the sample thickness $L$, and take the average to reduce the fluctuations. The number of realizations can vary from dozens to thousand, depending on the number of scatterers, and it is increased until one sees no significant changes in the transmission curves. We focus on densities $ \rho \leq 0.01 k_1^3 $ and optical thicknesses $b<1$ (see Appendix A), where multiple scattering orders are already required to describe coherent light scattering by cold atoms \cite{piovella2014}. Our goal is to show that EIT and applications are sensitive to coherent light-mediated interactions even for optically dilute regimes.

\begin{figure}[t]
\begin{centering}
\includegraphics[width=1.0\columnwidth]{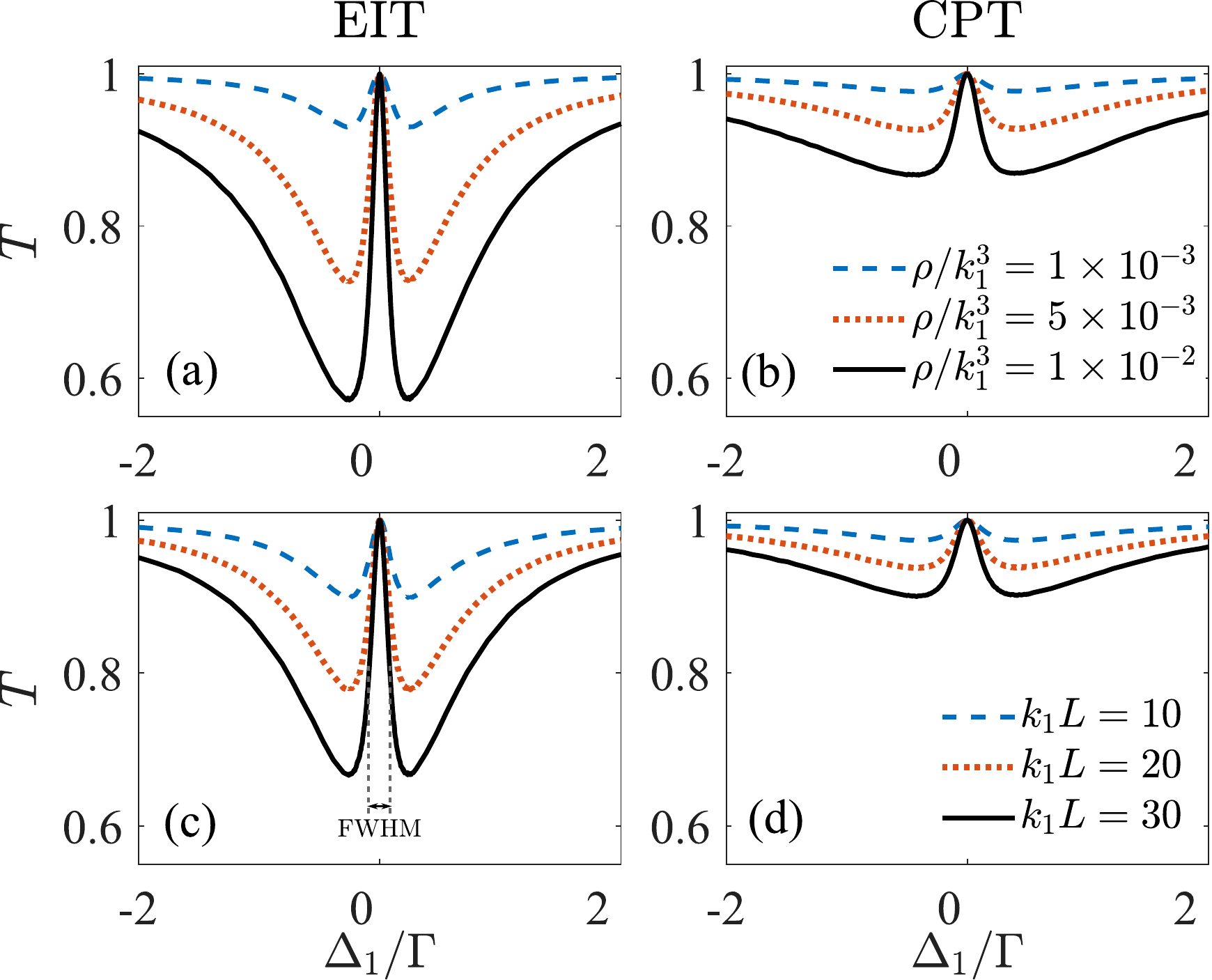}
\par\end{centering}
\caption{Transmission spectrum as a function of the probe field detuning. (a) and (c) were calculated in the EIT regime ($\Omega_2=0.5 \Gamma \gg\Omega_1=0.1\Gamma$), while (b) and (d) were obtained for the CPT regime ($\Omega_2=\Omega_1=0.5\Gamma$). In (a) and (b) we see the changes in the transmission spectrum for different values of the atomic density $\rho$, for a fixed cylinder thickness $k_1L=40$. While for (c) and (d) we vary $k_1L$ for a fixed density $\rho/k_1^3=0.01$. In panels (a) and (b),  the number of atoms $N$ ranged from $314$ to $3140$ and, in panels (c) and (d), from $785$ to $2335$.  For all plots we set $\Delta_2=0$ and $k_1R=50$. }
\label{fig:2}
\end{figure}

Fig. \ref{fig:2} displays the transmission spectrum as a function of the sample density and thickness, in the limits where the scalar model remains a good approximation ($\rho \lesssim 0.01k_1^3$) \cite{skipetrov2014}. In particular, Figs. \ref{fig:2}(a) and \ref{fig:2}(b) were obtained for $\Omega_2\gg\Omega_1$, a regime usually named as the ``EIT regime'', while (c) and (d) for $\Omega_1\sim\Omega_2$, the ``CPT regime'' \cite{ImamogluRev2005}. Note that the transparency at the resonance line ($\Delta_1=0$) remains unchanged for EIT and CPT regimes, so light-mediated interactions are not able to reduce the transparency maximum as Rydberg interactions do. \cite{adams2008,adams2010,fleischhauer2011,zhang2018Rydberg}. Such a difference arises from the very nature of the interactions: Rydberg interactions do not depend on how the atoms scatter light, whereas optical interactions totally disappear when the system reaches a dark state~\cite{ImamogluRev2005}. 

\begin{figure}[htpb]
\begin{centering}
\includegraphics[width=1.0\columnwidth]{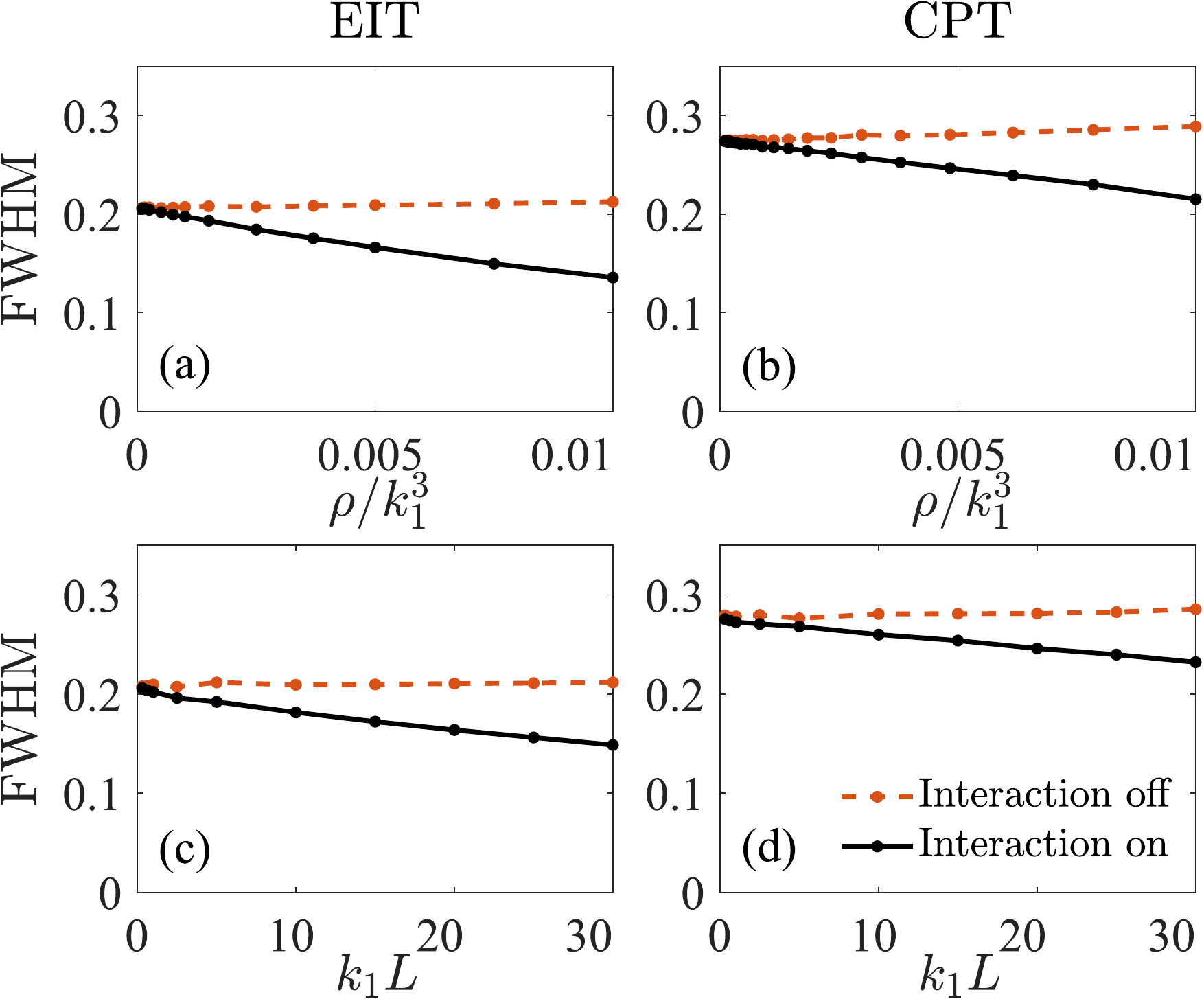}
\par\end{centering}
\caption{FWHM as a function of $\rho$ and $L$ for models with and without interacting terms. The full black curves are obtained by solving the full system of Equations \eqref{<sig11>}-\eqref{<sig12>}, whereas dashed orange curves by turning off all dipole interactions. (a) and (c) were calculated in the EIT regime ($\Omega_2=0.5 \Gamma \gg\Omega_1=0.1\Gamma$), while (b) and (d) were obtained for the CPT regime ($\Omega_2=\Omega_1=0.5\Gamma$). In (a) and (b) we see the changes in the FWHM by varying the atomic density $\rho$, for a fixed cylinder thickness $k_1L=40$, while for (c) and (d) we vary $k_1L$ for a fixed density $\rho/k_1^3=0.01$. For all plots we set $\Delta_2=0$ and $k_1R=50$, and the maximum number of atoms in the cloud in the simulations was $N=3142$}
\label{fig:3}
\end{figure}

Outside the resonance line, but still within the transparency window, we show that the FWHM is affected by collective effects for a not-so-low densities ($\rho  < 0.01 k_1^3$), a regime where recent works sought high efficiencies in quantum memories \cite{hsiao2018,vernaz2018,wang2019}. In Fig. \ref{fig:3}, we exhibit the FWHM for EIT and CPT transparency windows, as a function of the sample density $\rho$ and thickness $L$. We confront the results obtained from the full system of equations \eqref{<sig11>}-\eqref{<sig12>}, where the multiple scattering of light is preserved, with those of totally independent atoms ($G_{n}^{jl}=0$). The latter model predicts only single-scattering events, with no communication between the dipoles. Nevertheless, the scattered fields from each individual atom still interfere, thus being equivalent to highly rarefied atomic sample~\cite{bachelard2014,courteille2016}. As can be clearly noted in Figs. \ref{fig:3} (a-d), both models converge for very small densities and sample thickness (single-scattering regime), where weak incident fields ($\Omega_1,\Omega_2 \ll \Gamma_1,\Gamma_2$) leads to the limit $\textrm{FWHM}\propto \left(\Omega_1^2+\Omega_2^2\right)/\Gamma$ \cite{Gray78}. Yet, as the density increases for a fixed thickness (and vice-versa), a substantial disagreement appears between both models. The single-scattering model predicts a slow linear increment of EIT and CPT FWHM, while a practically linear narrowing of the transparency window is predicted by the complete interacting model even over the range where atomic clouds are still considered dilute ($\rho \lesssim 0.01 k_1^3$). In particular, Figs. \ref{fig:3}(a) and \ref{fig:3}(c), obtained for the EIT regime, show narrowings of $34.2\%$ and $27.8\%$, respectively, at the largest point of the horizontal axis. In the CPT regime (Figs. \ref{fig:3}(b) and \ref{fig:3}(d)), the narrowings are considerable smaller: $21.5\%$ and $17.0\%$, respectively. Our results point out that higher-order scattering events are relevant for the calculation of the narrowing of the transparency window \cite{piovella2014}.

We would like to highlight that narrowing of the EIT FWHM have been already estimated empirically from Beer-Lambert’s law \cite{lukin1997spectroscopy,ImamogluRev2005}, where many physical processes, as for instance, Doppler effect and collisions, contribute to this spectral narrowing. In these previous works, one deduces the scaling of the FWHM with the density from a naive extrapolation of the susceptibility for an ideal EIT non-interacting medium, where a Gaussian ansatz for the near-resonance transmittivity is considered \cite{lukin1997spectroscopy,ImamogluRev2005}. Our microscopic model, instead, predicts by first-principles the narrowing of the transparency window  provided by multiple-scattering effects (non-ideal medium), with no empirical assumptions for the transmission spectrum.

In the transmission profile, we have detected an asymmetry between positive and negative detunings (a little higher transmission for positive detunings). Such asymmetric behavior has been observed since the first experimental realization of the EIT phenomenon \cite{boller1991} and was wrongly attributed to several different effects, as Fano Interference and noninterfering photoionization channels \cite{boller1991}. However, the origin of such asymmetry relies on the interference that emerges between incident and scattered fields, whose mathematical term is proportional to the detuning $\Delta_1$ \cite{Jennewein2016}.  This feature is not unique to three-level systems. 

\begin{figure}[t]
\begin{centering}
\includegraphics[width=1.0\columnwidth]{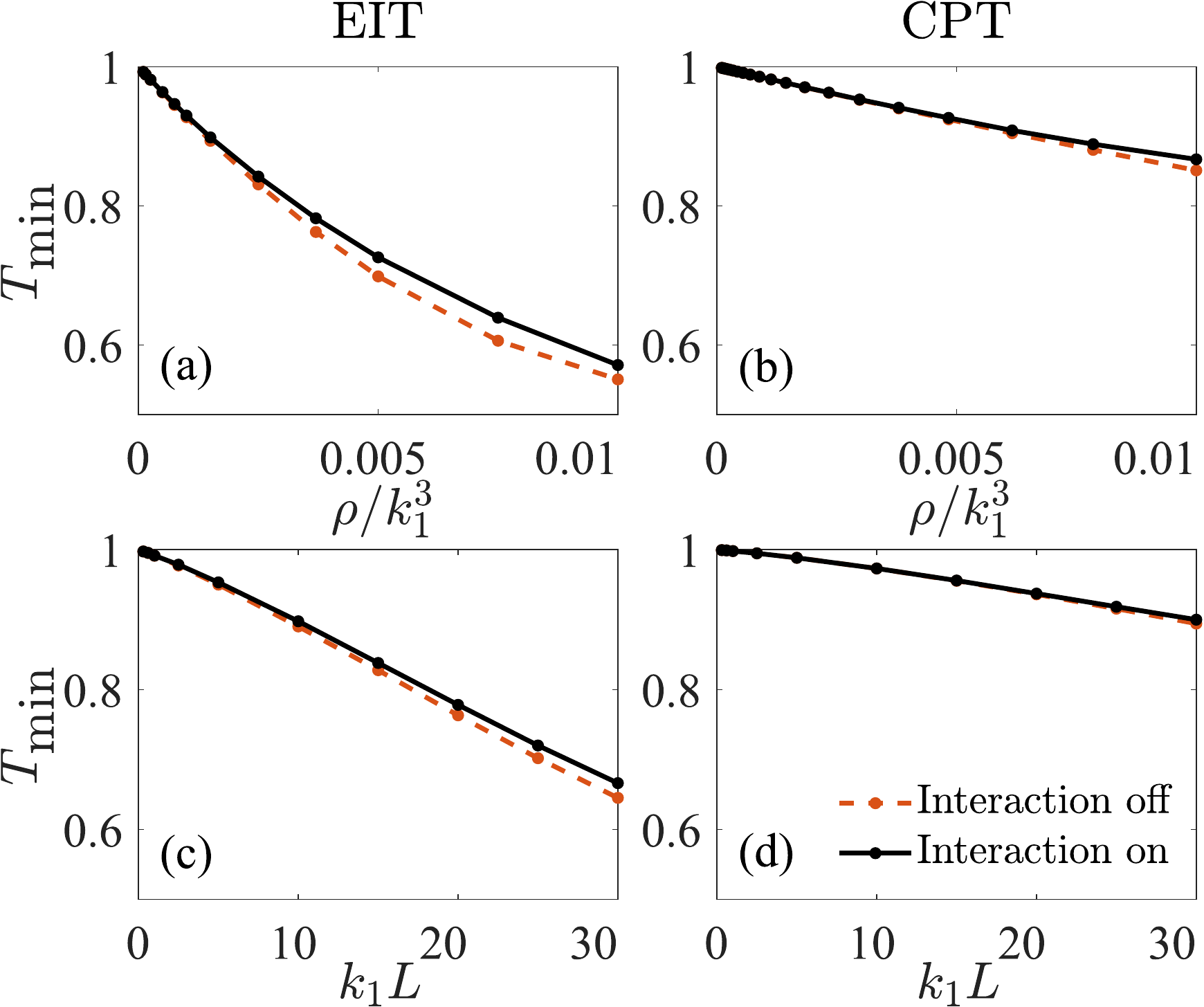}
\par\end{centering}
\caption{Minimum value of transmission as a function of $\rho$ and $L$. The full black curves are obtained by solving the full system of dynamical equations, whereas dashed orange curves by turning off all dipole interactions. (a) and (c) were calculated in the EIT regime ($\Omega_2=0.5 \Gamma \gg\Omega_1=0.1\Gamma$), while (b) and (d) were obtained for the CPT regime ($\Omega_2=\Omega_1=0.5\Gamma$). In (a) and (b) we see the changes in the FWHM by varying the atomic density $\rho$, for a fixed cylinder thickness $k_1L=40$, while for (c) and (d) we vary $k_1L$ for a fixed density $\rho/k_1^3=0.01$. For all plots we set $\Delta_2=0$ and $k_1R=50$, and the maximum number of atoms in the cloud in the simulations was $N=3142$. As a result of the asymmetry discussed in the text, we obtain different values of $T_{\textrm{min}}$ for opposite detunings, around $\Delta_1=\pm \sqrt{\Omega_1^2+\Omega_2^2}/2$, so $T_{\textrm{min}}$ refers to the lowest value between the two.}
\label{fig:4}
\end{figure}

Looking at the valleys of the transmission curves, we can see that minimum transmission $T_{\textrm{min}}$ (around $\Delta_1=\pm \sqrt{\Omega_1^2+\Omega_2^2}/2$) gets drastically reduced for increasing $\rho$ and $L$. Figs. \ref{fig:4}(a) and \ref{fig:4}(c), obtained for the EIT regime, show a reduction of $42.9\%$ and $33.4\%$ in the minimum transmission, respectively, for the last points of the  horizontal axis. Whereas in Figs. \ref{fig:4}(b) and \ref{fig:4}(d), obtained for the CPT regime, show a smaller variation: $13,3\%$ and $10\%$, respectively. We infer that $T_{\textrm{min}}$ is weakly affected by atomic interactions since the single-scattering model describes qualitatively well the reduction of the minimum transmission, over the regime where FWHW is incorrectly described by the same model. In other words, multiple scattering affects the FWHW even for not-so-high densities. The valleys of the transmission curves are actually modified for independent atoms when varying macroscopic parameters ($\rho$,$L$) because that the incident intensity $I_0$ is kept constant over the detection area $\mathcal{A}$ at the same time that the number of absorbers changes. 

\section{Collective STIRAP}
Now let us study how light-mediated interactions affect a coherent population transfer between the two ground states $\vert 1 \rangle \rightarrow \vert 2 \rangle$ via STIRAP \cite{Bergmann2019}, a key process for many quantum information applications, e.g., quantum memories \cite{ImamogluRev2005}.  To this goal, any slight probability of finding atoms in  state $\vert 1 \rangle$ is deleterious for the process. Therefore, taking care of precision, henceforward we compare the predictions from scalar and full vectorial models. Initializing the system with all atoms in the ground state $\vert 1\rangle$, we consider both fields varying in time as
\begin{eqnarray}
\Omega_{1}\left(t\right) & = & \Omega_{\textrm{max}}\left\{ \theta\left(t-t_{f}\right)\right.\nonumber \\
 &  & \left.+\sin\left(\frac{\pi t}{2t_{r}}\right)\left[\theta\left(t-t_{0}\right)-\theta\left(t-t_{f}\right)\right]\right\} ,\\
\Omega_{2}\left(t\right) & = & \Omega_{\textrm{max}}\left\{ 1-\theta\left(t-t_{0}\right)\right.\nonumber \\
 &  & \left.+\cos\left(\frac{\pi t}{2t_{r}}\right)\left[\theta\left(t-t_{0}\right)-\theta\left(t-t_{f}\right)\right]\right\} ,
\end{eqnarray}
where $\Omega_{\textrm{max}}$ represents the maximum value of the Rabi frequency of the fields, $t_0$ is the STIRAP process starting time, $t_r$ is the time it takes for the sine (cosine) to reach its maximum (minimum), $t_f=t_0+t_r$ is the instant where the variations in the fields end, and $\theta\left(x\right)$ is the Heaviside step function. The adiabaticity criterion is then fulfilled when $\Omega_{1},\Omega_{2} \gg \pi/2t_r$, and the population for independent atoms should be totally transferred coherently to the ground state $\vert 2 \rangle$ without ever populating the leaky excited state $\vert 3 \rangle$. In the previous section, we have studied the system's optical response in the steady state for a fixed ratio $\Omega_1/\Omega_2$, which lead the system to the stationary dark state $\vert1\rangle$. In the STIRAP, however, this ratio changes from $0$ to $\infty$. Now we initially prepare the system in $\vert1\rangle$, with the typical EIT configuration $\Omega_1\gg\Omega_2$, and vary the parameters until we reach $\Omega_1\ll\Omega_2$. This is the same regime as before but now with swapped roles for the two fields. The adiabatic theorem \cite{sakurai,griffiths} tells us that if the process is adiabatic, the system will now be at the new dark state $\vert2\rangle$.
\begin{figure}[ht]
\begin{centering}
\includegraphics[width=1.0\columnwidth]{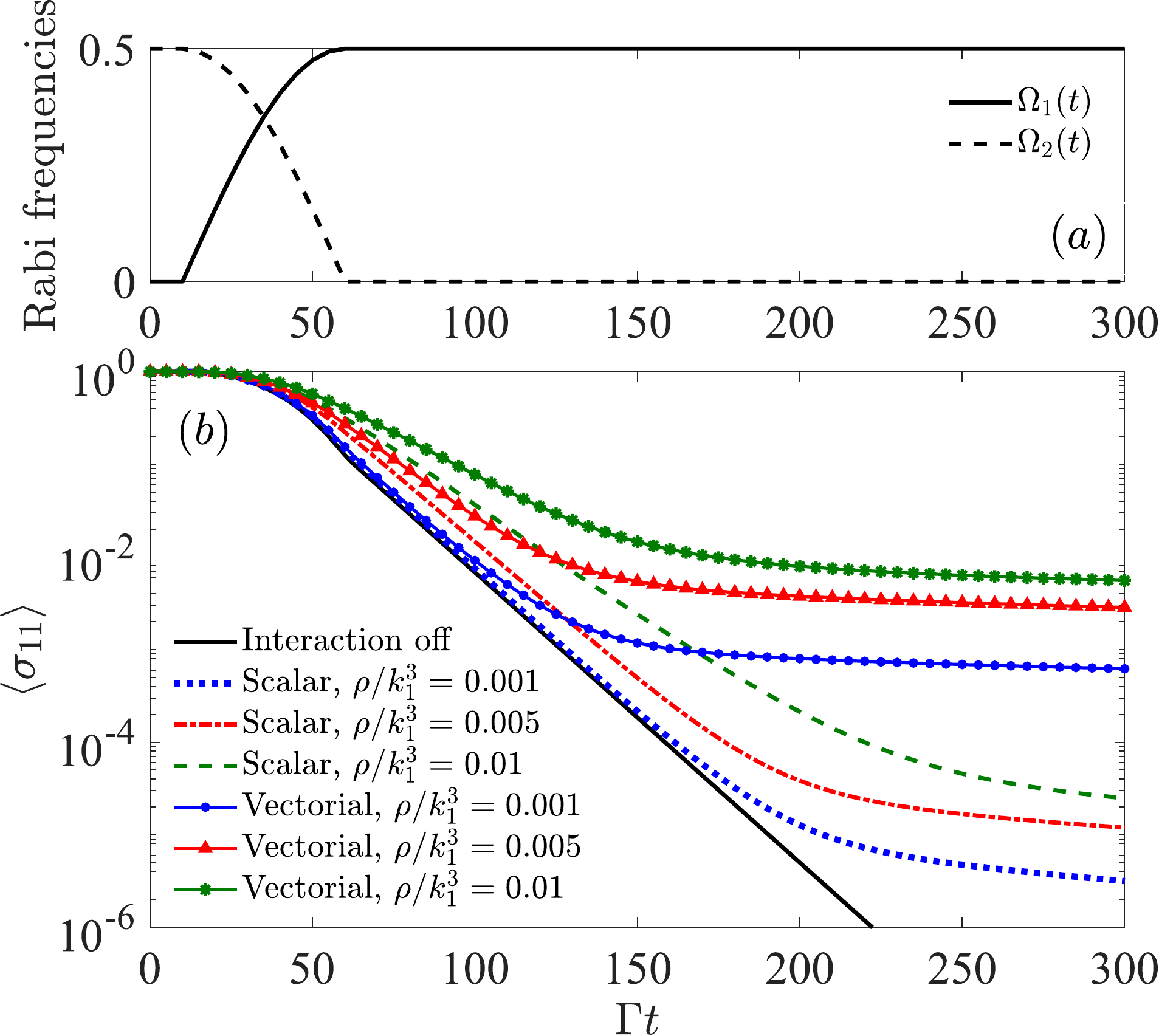}
\par\end{centering}
\caption{STIRAP process in a cold cloud of three-level atoms. Panel (a) shows how the Rabi frequencies of the probe and control field change in time, starting in a condition where $\Omega_1 \ll \Omega_2$ and adiabatically reaching a regime where $\Omega_1 \gg \Omega_2$. Panel (b) shows how the average ground state population of the state $\vert 1 \rangle$  behaves for different densities. For all plots we set $\Omega_{\textrm{max}}=0.5\Gamma$, $t_0=10\Gamma$, $t_r=60\Gamma$, $\Delta_1=\Delta_2=0$, $k_1L=60$ and $k_1R=40$. The number of atoms $N$ ranges from $1005$ to $3015$.}
\label{fig:5}
\end{figure}

 Panel (a) of Figure \ref{fig:5} shows the time variation of both fields, whereas painel (b) shows the average population $\langle \sigma_{11} \rangle$ of the ground state $\vert 1 \rangle$ throughout the process, for different values of density. Note that, after the STIRAP is over, the average population probability  $\langle \sigma_{11} \rangle$ remains for both scalar and vectorial models over much longer times than the noninteracting prediction. In particular, the vectorial model (represented by markers) shows a deviation around $1\%$ ($\langle \sigma_{11} \rangle \approx10^{-2}$) for $\rho/k_1^3 = 0.01$, two orders of magnitude higher than that from scalar model ($\langle \sigma_{11} \rangle \approx10^{-5}$). One could argue that such a error is not relevant, however, since the long-range interactions cannot be eliminated, they pose a fundamental limit for STIRAP based applications. For instance, from the results showed in Figure \ref{fig:5}, a qualitative estimation of the efficiency loss in the writing process of a quantum memory can be made. If the long-range interactions reduce the capacity of storing a photonic state in the atomic basis with high fidelity, in a quantum algorithm where the memory has to be accessed thousands of times \cite{nielsen2002quantum}, these errors will propagate and it is expected to drastically reduce the practical overall information retrieval efficiency of the quantum memory device. Our result points out that the efficiency can be even worse for increasing densities.

\section{Conclusion}
In conclusion, we have derived a model that describes the light scattering by a cold ensemble of three-level atoms. In the scalar regime, we were able to investigate how light-mediated long-range interactions influence EIT and CPT phenomena, by simulating the light transmission spectrum. This analysis demonstrates that optical dipole-dipole interactions considerably narrow the transparency window for sufficient dense and large atomic clouds, which can be useful for applications such as high-resolution spectroscopy. 
We have demonstrated that collective scattering also modifies a STIRAP process, showing that it spoils the population transfer between two atomic states. Although a propagating pulse model is required to infer a quantitative influence of these interactions on quantum memories, our microscopic analysis recreates the basic writing process of such devices and poses fundamental limitations to quantum memories.

Finally, we believe that our coupled-dipole model for three-level atoms, with control and probe fields, is useful for the study of many other situations. For example, it can be employed to investigate how nonlinear-optics effects \cite{ImamogluRev2005,binninger2019} are modified by collective scattering of light, as well as to understand if probe intensity profiles in space can be controlled by a control field. We can also investigate the modification of the efficiency for writing and generation of single photons in three-level systems via DLCZ protocol \cite{duan2001} for increasingly atomic densities.

\acknowledgments
This work was supported by the São Paulo Research Foundation (FAPESP) through grants No. 2019/11999-5 and No. 2017/13250-6, by the National Council for Scientific and Technological Development (CNPq) Grants No. 307077/2018-7 and 141247/2018-5. This work is also part of the Brazilian National Institute of Science and Technology for Quantum Information (INCT-IQ/CNPq) Grant No. 465469/2014-0. We thank R. Bachelard for the helpful discussions.

\appendix

\section{Scattering cross section and optical depth for a $\Lambda$ three-level atom}

In order to estimate the optical thickness for three-level atoms, we consider the main dynamical equations Eqs. \eqref{<sig11>}-\eqref{<sig12>} for the particular case of a single atom:
\begin{eqnarray}
\frac{d\left\langle \hat{\sigma}_{nn}\right\rangle }{dt} & = & \frac{\Gamma_{n}}{2}\left\langle \hat{\sigma}_{33}\right\rangle -\frac{i}{2}\Omega_{n}\left(\left\langle \hat{\sigma}_{n3}\right\rangle -\left\langle \hat{\sigma}_{3n}\right\rangle \right), \label{A1}\\
\frac{d\left\langle \hat{\sigma}_{12}\right\rangle }{dt} & = & -i\left(\Delta_{1}-\Delta_{2}\right)\left\langle \hat{\sigma}_{12}\right\rangle \nonumber \\
 &  & +\frac{i}{2}\Omega_{1}\left\langle \hat{\sigma}_{32}\right\rangle -\frac{i}{2}\Omega_{2}\left\langle \hat{\sigma}_{13}\right\rangle,\label{A2}\\
\frac{d\left\langle \hat{\sigma}_{n3}\right\rangle }{dt} & = & -\left[\frac{\left(\Gamma_{1}+\Gamma_{2}\right)}{2}+i\Delta_{n}\right]\left\langle \hat{\sigma}_{n3}\right\rangle \nonumber \\
 &  & -\frac{i}{2}\Omega_{n}\left(\left\langle \hat{\sigma}_{nn}\right\rangle -\left\langle \hat{\sigma}_{33}\right\rangle \right)-\frac{i}{2}\Omega_{m}\left\langle \hat{\sigma}_{12}\right\rangle,\label{A3} 
\end{eqnarray}
for $m,n=1,2$, with $m\neq n$. In the steady state, the expectation value of the excited state population is given by
\begin{widetext}
\begin{equation}
\langle \sigma_{33} \rangle_{ss} = \frac{4 \Gamma  \Delta _1^2 \Omega _1^2 \Omega _2^2}{\Gamma _2 \Omega _1^2 \left(4 \Gamma ^2 \Delta _1^2+\left(\Omega _1^2+\Omega _2^2\right){}^2\right)+\Omega _2^2 \left(\Gamma _1 \left(4 \Delta _1^2 \left(\Gamma ^2-2 \Omega _2^2\right)+16 \Delta _1^4+\left(\Omega _1^2+\Omega _2^2\right){}^2\right)+8 \Gamma  \Delta _1^2 \Omega _1^2\right)}.\label{sig33ss}
\end{equation}
\end{widetext}
The expression above allows us to obtain the scattering cross section $\sigma_{sc}=P_{sc}/I_{0}$, where $P_{sc}$ represents the scattered power, and $I_{0}\propto\left(\Omega_{1}/d_{1}\right)^{2}$ the incident field intensity. 

In order to obtain $P_{sc}$, we consider the scalar scattered field \eqref{E} in the far-field approximation:
\begin{equation}
E_{sc}^{(far)}( r,\hat{\mathbf{k}})\approx-\frac{\Gamma_{1}}{2d_{1}}\frac{e^{ik_{1}r}}{k_{1}r}\hat{\sigma}_{13}e^{-\hat{\mathbf{k}}.\mathbf{r}_{0}},
\end{equation}
where $\mathbf{r}_0$ is the position of the atom, and $\hat{\mathbf{k}}$ a unitary vector of observation in spherical coordinates. Since the scattered intensity is proportional to $I_{sc} \propto \langle E_{sc}^{\dagger}E_{sc} \rangle$, we then obtain
\begin{equation}
I_{sc}\left(r\right)\propto\left(\frac{\Gamma_{1}}{2d_{1}k_{1}r}\right)^{2}\left\langle \hat{\sigma}_{33}\right\rangle _{ss}.
\end{equation}
The integration of this intensity over an spherical shell results in the scattered power
\begin{eqnarray}
P_{sc} & \propto & \left(\frac{\Gamma_{1}}{2d_{1}k_{1}}\right)^{2}\left\langle \hat{\sigma}_{33}\right\rangle_{ss} \int_{0}^{2\pi}\int_{0}^{\pi}\sin(\theta)d\theta d\phi\\
\nonumber\\
 & \propto & \pi\left(\frac{\Gamma_{1}}{d_{1}k_{1}}\right)^{2}\left\langle \hat{\sigma}_{33}\right\rangle_{ss}.
\end{eqnarray}
Consequently, the scattering cross section for this three-level $\Lambda$ system can be expressed as
\begin{equation}
    \sigma_{sc}=\pi\left(\frac{\Gamma_{1}}{k_{1}\Omega_{1}}\right)^{2}\left\langle \hat{\sigma}_{33}\right\rangle_{ss},
\end{equation}
with $\left\langle \hat{\sigma}_{33}\right\rangle_{ss}$ given by Eq. \eqref{sig33ss}. Finally, we can calculate the optical thickness for this system integrating the density over the cylinder propagation direction \cite{guerin2017,kaiser2014}
\begin{eqnarray}
b&=&\sigma_{sc}\int_{-L/2}^{L/2}\rho\left(0,0,z\right)dz\\&=&\sigma_{sc}\rho L\\&=&\left(\frac{\Gamma_{1}}{\Omega_{1}}\right)^{2}\left\langle \hat{\sigma}_{33}\right\rangle _{ss}\frac{N}{k_{1}^{2}R^{2}},
\end{eqnarray}
where we used the fact that the average density $\rho=N/\pi R^2 L$ is constant over space \cite{akkermansbook,shengbook,guerin2017}. Therefore, when varying $L$ for a fixed homogeneous density, we are changing the optical thickness. Around the FWHM ($\Delta_{1}=0.125\Gamma$),  we obtain $b\approx 0.36$ for $\rho = 0.01 k_1^3$ and $ k_1 L = 40$, in the EIT regime: $\Gamma_{1}/\Gamma=\Gamma_{2}/\Gamma=0.5$, $\Omega_{1}=0.1\Gamma$ and $\Omega_{2}=0.5\Gamma$.

\bibliographystyle{apsrev4-1}
\bibliography{ref.bib}

\end{document}